\documentclass[useAMS,usenatbib]{mn2e}
\usepackage{graphicx}
\usepackage[dvips]{color}

\title[$\gamma$-rays from LLAGNs]{Gamma-rays as a diagnostic of the origin of core radiation in low-luminosity active galactic nuclei}
\author[H. Takami]{Hajime Takami$^1$
\thanks{E-mail:takami@mpp.mpg.de} \\
$^{1}$Max Planck Institute for Physics, F$\ddot{\rm o}$hringer Ring 6, 80805 Munich, Germany}
\begin{document}

\date{Submitted \today}

\pagerange{\pageref{firstpage}--\pageref{lastpage}} \pubyear{2010}

\maketitle

\label{firstpage}

\begin{abstract}
The respective contribution of disk and jet components to the total emission in low luminosity active galactic nuclei (LLAGNs) is an open question. This paper suggests that $\gamma$-rays emitted from electrons accelerated in jets could be a direct diagnostic tool for a jet component to the total emission. We demonstrate $\gamma$-ray flux from jets based on a synchrotron self-compton (SSC) model on the assumption that radio and X-rays are dominantly produced from jets in the case of a high state of a nearby LLAGN, NGC 4278. We also survey parameter space in the model. Observational properties of LLAGNs in radio and X-ray bands allow to constrain physical parameters in an emission region. The size of the emission region $R$ is limited to $10^{16}$ cm $\leq R \leq 10^{17.5}$ cm if the observed tight correlation between radio and X-ray emission originates from the same jet component. If the beaming factor of the emission region is close to the observed parsec scale jet of NGC 4278 and $R \sim 10^{16}$ cm, the $\gamma$-rays may be detected by Cherenkov Telescope Array, and the jet domination can be tested in the near future. 
\end{abstract}


\section{Introduction} \label{introduction}

Active galactic nuclei (AGNs) are powerful objects in the Universe, which are thought to be powered by gravitational energy through the accretion of surrounding gas onto supermassive black holes (SMBHs) located at the center of the accretion systems. Classically, the accretion paradigm has been applied to extremely powerful objects like quasars and radio galaxies for extragalactic objects. The Eddington ratio $L_{\rm bol}/L_{\rm Edd}$ of these objects, where $L_{\rm bol}$ and $L_{\rm Edd}$ are bolometric luminosity and Eddington luminosity, respectively, is $\sim 10\%$ and thereby geometrically thin, optically thick disk models, so-called standard disks \citep{Shakura1973AA24p337}, have been successfully favored. Recently, the idea that all the galaxies have SMBHs at their centers has been commonplace \citep{Kormendy1995ARAA33p581,1998AJ115p2285}. Optical spectroscopic surveys have revealed that a significant fraction of nearby galaxies has also active nuclei, but much less luminous than the powerful objects ($< 10^{40}$ erg$^{-1}$ in nuclear H$\alpha$ luminosiy), called low luminosity AGNs (LLAGNs) \citep{Ho1997ApJS112p315}. LLAGNs can be furthermore spectroscopically classified into Low Ionization Nuclear Emission Region nuclei (LINERs), Seyfert galaxies, and transition objects. Their Eddington ratios are much lower than the canonical value ($\sim 10\%$) appropriate for standard disks and could reach $\sim 10^{-8}$ in some cases (e.g., \citet{Ho2009ApJ699p626}). In addition to the low radiative efficiency, correlation among radio luminosity, X-ray luminosity, and the mass of nuclear black holes \citep{Merloni2003MNRAS345p1057,Falcke2004AA414p895} and the weak feature of the big blue bump \citep{Ho2008ARAA46p475} indicate optically thin, radiatively inefficient accretion flows (RIAFs) in LLAGNs.

Advection dominated accretion flow (ADAF) models \citep{Narayan1994ApJ428L13,Narayan1995ApJ444p231,Abramowicz1995ApJ438p37}, are a kind of RIAFs and have been widely discussed to explain the spectral energy distribution (SED) of LLAGNs. ADAF models successfully explained most parts of the SED of Sagittarius A$^*$ which is a SMBH located at the center of our Galaxy \citep{Narayan1995Natur374p623,Manmoto1997ApJ489p791,Narayan1998ApJ492p554,Mahadevan1998Natur394p651}.

ADAF models have been applied to nearby LLAGNs. In order to investigate the origin of radiation from LLAGN core, radio and X-ray bands are often adopted to avoid possible contamination from stellar and dust emission. Follow-up observations of optically selected LLAGNs in radio bands have detected emission dominated with a compact core morphology \citep{Ho2001ApJS133p77,Nagar2002AA392p53}. Despite their low luminosity, a larger fraction of LLAGNs are radio-loud \citep{Ho2001ApJ555p650,Terashima2003ApJ583p145}. \citet{Doi2005MNRAS363p692} detected a spectral bump in high-frequency radio ($\sim$ submillimeter) bands which is a spectral feature of an ADAF component. However, observations have revealed in radio bands (1-10 GHz) that ADAF components are not enough to reproduce the total radio emission, which indicate the contribution of jets \citep{Ulvestad2001ApJ562L133,Anderson2004ApJ603p42,Wu2005ApJ621p130}. These observations imply the coexistence of both disk and jet components in emission from LLAGNs.

The respective contribution of disk and jet components from radio to X-ray bands is an open problem. \citet{Merloni2003MNRAS345p1057} found correlation among radio luminosity, X-ray luminosity, and black hole mass and showed that an ADAF model can reproduce observational data better than a radiation model dominated by jets by comparing the data with theoretical models. On the other hand, \citet{Falcke2004AA414p895} suggested that radiation is dominated by non-thermal emission from jets by a similar analysis. The correlation between radio and X-ray luminosities has also implied the common non-thermal origin from jets for radio-loud LLAGNs \citep{Balmaverde2006AA447p97,Panessa2007AA467p519}. Note that \citet{Merloni2003MNRAS345p1057} also mentioned that considering cooling processes of electrons can improve the reproducibility of the data in the jet model and synchrotron radiation from jets can be responsible up to X-rays. Furthermore, \citet{Nemmen2010IAUS267p313} demonstrated the SED fittings of observed LLAGNs and showed that observed X-ray data can be reproduced by both a disk-dominated and jet-dominated radiation models without inconsistency at present.

This study suggests that $\gamma$-rays emitted from electrons accelerated in jets can be a direct diagnostic tool for a jet component in radio to X-ray bands. When X-rays are produced by synchrotron radiation of the electrons, the synchrotron photons are also upscattered by the electrons through inverse compton scattering (ICS) into $\gamma$-rays. This is so-called synchrotron self-compton (SSC) scenario \citep{Maraschi1992ApJ397L5,Bloom1996ApJ461p657}. This scenario is well established for the SED modelings of BL Lac objects (e.g., \citet{Anderhub2009ApJ705p1624,Aharonian2009ApJ696L150}). We focus on NGC 4278 as an example of LLAGNs and demonstrate expected $\gamma$-ray flux based on a SSC model on the assumption that non-thermal radiation from jets is dominated. We also survey parameter space in the model and constrain physical parameters required in jets.

NGC 4278 is a LLAGN with $z = 0.002$ and the distance of 16.7 Mpc \citep{Tonry2001ApJ546p681} for $H_0 = 70$ km s$^{-1}$ Mpc$^{-1}$, which is Hubble constant. This source is sometimes categorized into LINER or radio-loud LLAGNs. Recent observations in radio bands have revealed a two-sided relativistic parsec scale jet ($\beta \sim 0.75$) closely aligned to the line of sight ($2^{\circ} \leq \theta \leq 4^{\circ}$), where $\beta$ and $\theta$ are the velocity in the unit of speed of light and the viewing angle of the jet, respectively \citep{Giroletti2005ApJ622p178}. These observables lead to the relativistic beaming factor of the parsec scale jet of $2.6$. In addition, a monthly time scale variability by a factor of 3 to 5 has been observed in X-ray bands \citep{Younes2010AA517p33}. In general, the time scale of flux variability limits the size of emission region.

This paper is laid out as follows. In Section \ref{sec:model}, we describe a model of SED originating from electrons accelerated in relativistic jets. In Section \ref{sec:gamma}, we fit resultant SED to observed data below X-rays and estimate $\gamma$-ray flux. Moreover, combining observational results in radio and X-ray bands, we constrain the physical parameters of emission region. Finally, we make some discussions and summarize this study in Section \ref{sec:sum}.

\section{Model} \label{sec:model}

Electrons are accelerated in a jet via particle acceleration mechanisms, i.e., shock acceleration \citep{Blandford1987PhR154p1} and emit synchrotron radiation by interactions with magnetic field. The synchrotron photons are also upscattered by the electrons through ICS. For simplicity, this electron acceleration region is modeled as a magnetized spherical blob with the radius of $R$, which is moving with the Lorentz factor of $\Gamma$ in a jet. Once we inject accelerated electrons into the blob, we discuss the interactions and emission from the blob following a SSC model. In this section, we introduce a model of relativistic electron spectrum. The other details of our SSC model is described in appendix \ref{app:ssc}.

Although statistical particle acceleration mechanisms produce a power-law spectrum of electrons, the cooling processes of the electrons make a spectral break. Based on this picture and the approximation that electrons are continuously injected to the blob during the order of $R/c$, electrons accelerated in the blob are assumed to have a broken power-law spectrum, 
\begin{equation}
\frac{d n_e}{d\gamma} = n_0 \gamma^{-s_1} 
\left( 1 + \frac{\gamma}{\gamma_{\rm br}} \right)^{s_1 - s_2} 
~~~ \left( \gamma_{\rm min} < \gamma < \gamma_{\rm max} \right), 
\label{eq:sp}
\end{equation}
where $n_0$, $\gamma$ are the normalization factor of the number density of the electrons and the Lorentz factor of electrons, respectively. $\gamma_{\rm min}$, $\gamma_{\rm max}$, and $\gamma_{\rm br}$ are the minimum, the maximum of $\gamma$, and and the Lorentz factor at a spectral break, respectively. We set $\gamma_{\rm min} = 1$ throughout this paper. $s_1$ and $s_2$ are the spectral indices below and above $\gamma_{\rm br}$. For variable sources, an electron spectrum is expected to be time-dependent in reality, but the time profile of the electron injection is uncertain at present. A simple model is instantenious injection approximation discussed in literture (e.g., \citet{Dermer1993ApJ416p458,Dermer2002ApJ575p667}). On the other hand, when particle acceleration at shocks in a jet is considered, the shocks propagate over a size of the out-flowing plasma and can deposite energy continuously as also mentioned in \citet{Dermer1993ApJ416p458}. Thus, continuous injection approximation is also plausible in internal shock scenarios. It should be constrained by observations which injection approximation is closer to reality. In this paper, we adopt the continuous injection approximation and the electron spectrum described in Eq.\ref{eq:sp}.

$\gamma_{\rm max}$ can be estimated by comparing the time scale to accelerate electrons and the shortest energy-loss time scale. The former time scale is 
\begin{equation}
\tau_{\rm acc} = \theta_F \frac{r_g}{c} 
= 6 \times 10^2 \theta_{F,2} \gamma_7 {B_{-1}}^{-1} ~~~{\rm s}, 
\end{equation}
where $r_g$ and $c$ is the Larmor radius of an electron and the speed of light, respectively, and $\theta_{F,2} = \theta_F / 10^2$, $\gamma_7 = \gamma / 10^7$, $B_{-1} = B / 10^{-1}$ G. We take $\theta_F$ to be a constant, which is equivalent to assuming the diffusion coefficient for accelerated particles to be proportional to the Bohm diffusion coefficient. $\theta_F \geq 10$ is a fairly conservative value, but $\theta_F \sim 1$ is also possible for relativistic shocks \citep{Rachen1998PRD58p123005}.

The accelerated electrons suffer from energy-loss via synchrotron radiation and ICS. The former time scale is 
\begin{equation}
\tau_{\rm syn} = \frac{3 m_e c}{4 \sigma_T \gamma U_B} 
= 8 \times 10^3 {\gamma_7}^{-1} {B_{-1}}^{-2} ~~~{\rm s}, 
\end{equation}
where $m_e$, $\sigma_T$, and $U_B = B^2 / 8\pi$ are the electron mass, the cross-section of Thomson scattering, and the energy density of magnetic field, respectively. The time scale of ICS has a form similar to $\tau_{\rm syn}$ in the Thomson limit, 
\begin{equation}
\tau_{\rm ics} = \frac{3 m_e c}{4 \sigma_T \gamma U_{\rm rad}}, 
\end{equation}
where $U_{\rm rad}$ is the energy density of radiation. The shortest energy-loss time scale is determined by comparing $U_{\rm B}$ and $U_{\rm rad}$. If $U_{\rm rad}$ is larger than $U_{\rm B}$, $\gamma_{\rm max}$ and $\gamma_{\rm br}$ are estimated by adopting $U_{\rm rad}$. The flaring of blazars realize $U_{\rm rad} > U_{\rm syn}$ in many cases. However, $U_{\rm rad}$ cannot be estimated before the calculation of SED. Returning to a kinetic equation of electrons to determine electron distribution, the equation is non-linear because ICS energy-loss rate depends on the electron distribution. This non-linearity changes resultant $\gamma$-ray spectrum \citep{Schlickeiser2009MNRAS398p1483,Schlickeiser2010AA519A9,Zacharias2010AA524A31}. Since we do not know $U_{\rm rad}$ initially on the spectral assumption of Eq.\ref{eq:sp}, first of all, we calculate SEDs on the assumption of $U_{\rm B} > U_{\rm rad}$, which means that the shortest time scale is $\tau_{\rm syn}$, and then check $U_{\rm rad}$. Although this assumption will be satisfied in many cases, $U_{\rm B} < U_{\rm rad}$ could be realized when $R$ is small. Fortunately, such cases are not favored by several observational aspects as discussed in Section \ref{sec:gamma}. Consequently, $\gamma_{\rm max}$ can be estimated by $\tau_{\rm acc} = \tau_{\rm syn}$ under this assumption as 
\begin{equation}
\gamma_{\rm max} = \left( 
\frac{6 \pi e}{\theta_F \sigma_T B} \right)^{-1/2} 
= 4 \times 10^7 {\theta_{F,2}}^{-1/2} {B_{-1}}^{-1/2}. 
\label{eq:gmax} 
\end{equation}
Note that this $\gamma_{\rm max}$ always satisfies the Hillas criterion that $r_g$ should be smaller than $R$ \citep{Hillas1984ARAA22p425} in all the cases treated in this paper.

The cooling of electrons via synchrotron radiation and/or ICS makes a power-law index be steeper by one, $s_2 = s_1 + 1$ above characteristic energy $\gamma_{\rm br}$, assuming that accelerated electrons are continuously provided into the blob. Since the electrons lose energy during their staying in the blob, this characteristic energy can be estimated by $\tau_{\rm syn} < \tau_{\rm esc}$, where 
\begin{equation}
\tau_{\rm esc} = \frac{3R}{c} 
= 1 \times 10^6 R_{16} ~~~{\rm s}. 
\end{equation}
is the time scale to escape electrons from the blob. Here, we assume that electrons escape from the blob by the velocity of $c/3$, which corresponds to the velocity of downstream fluid in a shock system under the strong shock limit of a relativistic shock. Thus, 
\begin{equation}
\gamma_{\rm br} = \frac{2 \pi m_e c^2}{\sigma_T R B^2} 
= 8 \times 10^4 {R_{16}}^{-1} {B_{-1}}^{-2}. 
\label{eq:gbr} 
\end{equation}

We estimate the SEDs of a jet component of NGC 4278 in the overall energy range based on the SSC model under the electron spectrum modeled above. This model has five free parameters: $s_1 (s_2)$, $B$, $R$, $n_0$, and $\delta$, where $\delta = \left[ \Gamma ( 1 - \beta \cos \theta ) \right]^{-1}$ is so-called the beaming factor of the blob. For convenience, we change a parameter $n_0$ into the ratio of electron energy density in the blob to $U_B$, $\eta = U_e / U_B$, where 
\begin{equation}
U_e = m_e c^2 \int_{\gamma_{\rm min}}^{\gamma_{\rm max}} 
d\gamma \gamma \frac{d n_e}{d\gamma}. 
\end{equation}

\section{Expected $\gamma$-rays} \label{sec:gamma}

In order to estimate $\gamma$-ray flux by fitting a spectrum observed in radio to X-ray bands, we adopt data processed by \citet{Younes2010AA517p33}. A selection criterion of this data set is high angular resolution less than 10 arc-seconds to resolve radiation from the core of NGC 4278. The authors adopted the data of both Chandra and XMM-Newton obtained in different periods in X-ray bands and found a monthly time scale variability in which flux is changed by a factor of 3 to 5. In this paper, we focus on a high state observed by XMM-Newton due to following two reasons. The first reason is the existence of near-UV data simultaneously observed with the X-rays by an optical/UV monitor instrument on board XMM-Newton. Simultaneous data for variable sources are generally a powerful tool to unveil the origin of emission from the sources. The second reason is that a high state expects larger $\gamma$-ray flux, which allows us to diagnose the origin of core radiation more easily. Since XMM-Newton does not have angular resolution as well as Chandra, it could not resolve two X-ray point sources close to the core resolved by Chandra. The authors estimated X-ray components other than the core as only 2\% by using Chandra data. The data could be well fitted by a power-law spectrum without any thermal component and confirmed that a spectral index in X-ray bands is not contaminated. The photon index defined in the X-ray spectrum in the unit of photons cm$^{-2}$ s$^{-1}$ keV$^{-1}$ is $2.05 \pm 0.02$. Assuming that the X-rays are synchrotron radiation from high energy electrons above $\gamma_{\rm br}$, the spectral index of the electrons is $s_2 \sim 3.0$. The other optical data was processed from data observed by Hubble Space Telescope (HST), but they are not observed simultaneously. In radio bands, we adopt data observed by \citet{Nagar2001ApJ559L87},\citet{Giroletti2005ApJ622p178}, \citet{Jones1984ApJ276p480}, which were also used and listed in \citet{Younes2010AA517p33}. The spectral index $\alpha_{\rm rad}$ defined as $\nu L_{\nu} \propto \nu^{-\alpha_{\rm rad}}$ in radio bands is $\sim 0.5$, where $L_{\nu}$ is luminosity per unit frequency, which corresponds to $s_1 \sim 2.0$ because $s_1 = 2 \alpha_{\rm rad} + 1$ for synchrotron radiation. In order to produce SED in overall energy range, we fit the observed data in radio and X-ray bands by the model. The variability of radio emission is assumed to be small, although the radio data are not simultaneous data to X-rays.

\begin{figure}
\begin{center}
\includegraphics[width=0.98\linewidth]{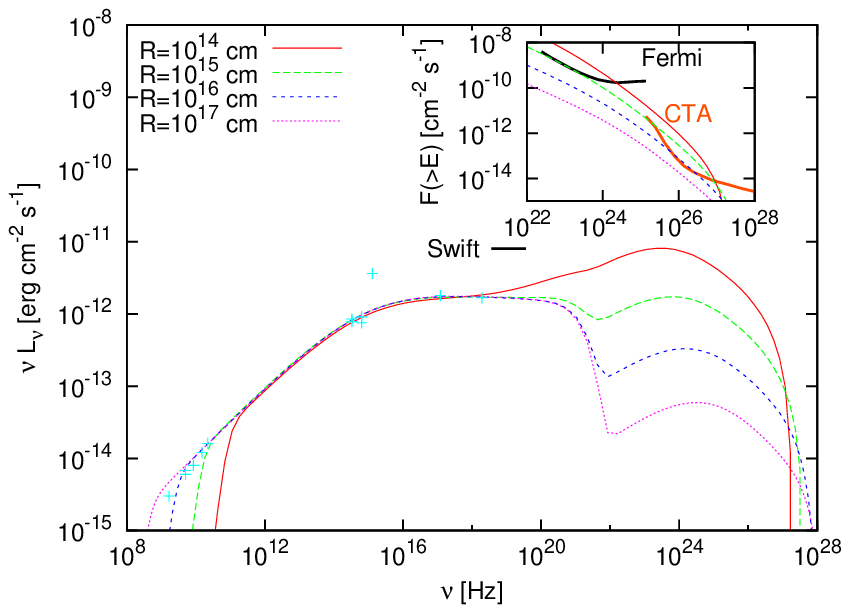} 
\caption{Spectral energy distributions on the assumptions of $\delta = 2.6$ 
and $\theta_F = 10^2$. The other parameters are listed in Table 
\ref{tab:parameters}. The sensitivity limit of Swift BAT 54 months 
sky survey \citep{Cusumano2010arXiv1009.0522} is also shown. 
In addition, the corresponding integral fluxes, $F(>E)$, are shown 
with the integral sensitivity curves of Fermi LAT (5$\sigma$, 1yr) 
\citep{Atwood2009ApJ697p1071} and CTA (5$\sigma$, 100hr) 
\citep{CTA2010arXiv1008.3703} in the small panel. 
The cases of $R = 10^{14}$ and $10^{15}$ cm are not favored 
because of synchrotron self-absorption (see text).}
\label{fig:sed1}
\end{center}
\end{figure}

\begin{table}
\begin{center}
\begin{tabular}{c|c|c} \hline
R [cm] & B [G] & $\eta$ \\ \hline \hline
$10^{14}$ & 3.72 & 140.0 \\
$10^{15}$ & 0.80 & 37.0 \\
$10^{16}$ & 0.165 & 10.0 \\
$10^{17}$ & 0.035 & 2.5 \\ \hline
\end{tabular}
\caption{Physical parameters adopted in Figs.\ref{fig:sed1} and \ref{fig:sed2}. 
$\delta = 2.6$ is assumed in these figures.}
\label{tab:parameters}
\end{center}
\end{table}

The size of an emission blob can be constrained by several discussions. The monthly time scale variability in X-ray bands can limit the blob size as 
\begin{equation}
R \leq \frac{\delta c \Delta t_{\rm obs}}{1 + z} 
= 3 \times 10^{17} \delta_3 \Delta t_{\rm obs,-1} ~~{\rm cm}, 
\end{equation}
where $\delta_3 = \delta / 3$ and $\Delta t_{\rm obs,-1} = \Delta t_{\rm obs} / 10^{-1}$ yr is the variability time scale. Although $\sim 1.5$ hours $\sim 10\%$ variability in X-ray flux is also reported in the term observed by XMM-Newton, the origin of this variability does not care here, but will be discussed afterward. On the other hand, assuming that X-rays are emitted from a blob in jets, its size is difficult to be smaller than the Schwarzschild radius of the central black hole, $r_{\rm sch}$. Since the black hole mass of NGC 4278 is $M_{\rm BH} \sim 3 \times 10^8 M_{\odot}$ \citep{Wang2003MNRAS340p793, Chiaberge2005ApJ625p716}, the blob size is lower-limited as 
\begin{equation}
R \geq r_{\rm sch} = \frac{2GM_{\rm BH}}{c^2} 
= 9 \times 10^{13} M_{\odot,8.5} ~~{\rm cm}, 
\end{equation}
where $M_{\odot,8.5} = M_{\rm BH} / 10^{8.5} M_{\odot}$, $G$ and $M_{\odot}$ are the gravitational constant and the solar mass, respectively. Thus, we consider $10^{14}$ cm $\leq R \leq$ $10^{17}$ cm below.

\begin{figure}
\begin{center}
\includegraphics[width=0.98\linewidth]{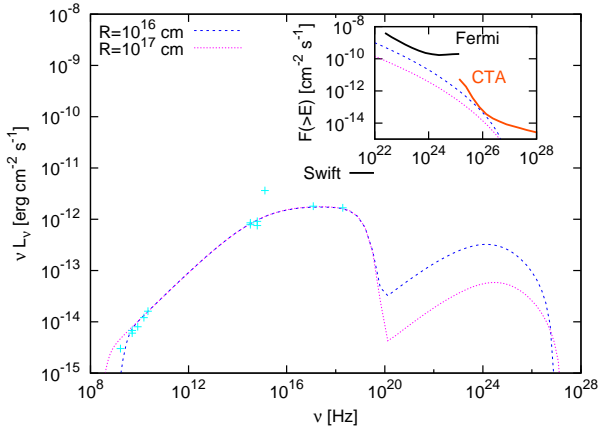} 
\caption{Same as Fig.\ref{fig:sed1}, but the cases of $\theta_F = 10^4$.}
\label{fig:sed2}
\end{center}
\end{figure}

Fig.\ref{fig:sed1} shows SEDs calculated on the assumptions of $\delta = 2.6$ and $\theta_F = 10^2$. The absorption of high-energy $\gamma$-rays by extragalactic background light is taken into account by using the model of \citet{Franceschini2008AA487p837}. The other parameters required for the spectral fits are listed in Table \ref{tab:parameters}. Note that $\delta = 2.6$ was derived from a radio image of a parsec scale jet \citep{Giroletti2005ApJ622p178}. $\delta$ could be larger at inner jets (discussed afterward) as radio observations have revealed mildly relativistic parsec scale jets in blazars contrary to the requirement of $\delta \geq 10$ to reproduce $\gamma$-ray flux \citep{Piner2004ApJ600p115}. The sensitivity limit of Swift Burst Alert Telescope (BAT) 54 months sky survey \citep{Cusumano2010arXiv1009.0522} is also shown. In the small panel inside the figure, integral fluxes corresponding to the SEDs with the integral sensitivity curve of Fermi Large Area Telescope (LAT) (5$\sigma$, 1yr) \citep{Atwood2009ApJ697p1071} and goal integral sensitivity of Cherenkov Telescope Array (CTA) (5$\sigma$, 100hr) \citep{CTA2010arXiv1008.3703}.

Fig.\ref{fig:sed1} demonstrates that a smaller blob predicts larger $\gamma$-ray flux. The case of a smaller blob requires larger number density of synchrotron photons, i.e., seed photons for ICS, to reproduce the observed flux up to X-rays, which leads to larger $\gamma$-ray flux. To realize the increase of synchrotron photons, a strength of magnetic field and/or the number of electrons in the blob should increases. However, only the number of electrons is essentially a free parameter, since the strength of magnetic field is constrained to reproduce the observed data as follows. In order to fit the radio and X-ray data, a spectral break of synchrotron radiation should be at $\nu_{\rm br} \sim 10^{15}$ Hz. Since the typical frequency of synchrotron photons generated from electrons with the energy of $\gamma$ is 
\begin{equation}
\nu_c = 0.29 \frac{\delta}{1 + z} \frac{3 \gamma^2 e B}{4 \pi m_e c} 
= 4 \times 10^5 \delta_3 {R_{16}}^{-2} {B_{-1}}^{-3} \gamma^2 ~~{\rm Hz}, 
\label{eq:typnu}
\end{equation}
the requirement of $\nu_{\rm br} \sim 10^{15}$ Hz means 
\begin{equation}
{R_{16}}^2 {B_{-1}}^3 \sim 7 \times 10^{-1} \delta, 
\label{eq:rbconst}
\end{equation}
from Eqs.\ref{eq:gbr} and \ref{eq:typnu}. The values of magnetic field listed in Table \ref{tab:parameters} roughly consistent with this estimation. In Table \ref{tab:parameters}, $\eta$ is larger for a smaller blob, which means that more electrons are required for a smaller blob with respect to the increase of magnetic field. Also, the electron energy density is dominated compared to the energy density of magnetic field for $\delta = 2.6$. This is a similar situation to the emission region of blazars.

However, the small size of the blob has several problems. Firstly, radiation energy density is dominated compared to the energy density of magnetic field in the blob in the cases of $R = 10^{14}$ and $R = 10^{15}$ cm. In these cases, $\gamma_{\rm max}$ and $\gamma_{\rm br}$ are determined by ICS, which are unlikely to Eqs.\ref{eq:gmax} and \ref{eq:gbr}. However, even including the effects of ICS, the model cannot reproduce the observed data self-consistently on the assumption of $\delta = 2.6$. $U_{\rm rad}$ much larger than $U_{\rm B}$ makes the spectral break frequency of synchrotron radiation be lower. In order to keep $\nu_{\rm br}$ to be $\sim 10^{15}$ Hz, $U_{\rm rad}$ should be reduced by decreasing $n_e$ or $B$. Both of these decreases the flux of synchrotron radiation. In the former case, $B$ is required to be larger to compensate the decrease of synchrotron radiation, but this leads to a break frequency lower than that calculated in Fig.\ref{fig:sed1}. Although $\eta$ needs to be larger to reproduce the observed data in the latter case, the total $U_{\rm rad}$ is not reduced because ICS flux is proportional to ${n_e}^2$. The second reason is another spectral break due to synchrotron self-absorption. A smaller $R$ produces the spectral break at a higher frequency, which is $\sim 10^{11}$ Hz and $\sim 10^{10}$ Hz for $R = 10^{14}$ and $R = 10^{15}$ cm, respectively. In these cases, radio emission from the blob is not dominated at $\sim 5$ GHz. For well-observed blazars, radio emission at 5 GHz is already optically thick and therefore radio emission from the blob is not dominated (e.g., \citet{Anderhub2009ApJ705p1624,Aharonian2009ApJ696L150}). It is not clear at present whether this feature is common in radio-loud LLAGNs. However, the tight correlation between radio luminosity at 5 GHz and X-ray luminosity implies that the origin of these radiations is the same \citep{Panessa2007AA467p519}. Based on this implication, $R \leq 10^{15.5}$ cm is not favored for $\delta = 2.6$.

The integral flux for $R = 10^{16}$ and $10^{17}$ cm are lower than the integral sensitivity curve of Fermi LAT. This is consistent with the fact that NGC 4278 is not listed in Fermi LAT 1yr source catalog \citep{Abdo2010ApJS188p405}. Unfortunately, even 10 years observations do not reach the prediction for $R = 10^{16}$ cm, assuming that the sensitivity curve of Fermi LAT is roughly scaled inversely proportional to the square root of the total observation time. The integral flux is also compared with the goal integral sensitivity of CTA. The integral flux for $R = 10^{16}$ cm is above the CTA sensitivity at $\sim 100$ GeV. Thus, CTA can test the jet domination of the total emission in this source in the case of $R \sim 10^{16}$ cm.

The simultaneous data in a near-UV band is not consistent with the synchrotron component of the SEDs, while the (non-simultaneous) data in optical bands are well fitted by the synchrotron component. Since an one zone synchrotron model cannot satisfy both this near-UV point and a flat spectrum in X-ray bands at the same time, this near-UV radiation is thought to be the contribution from an accretion disk. In ADAF models, the ICS component of high-temperature thermal electrons in the disk can make a spectral peak at this band　(e.g., \citet{Manmoto1997ApJ489p791,Nemmen2010IAUS267p313}). The synchrotron component is not always dominated in all the bands below X-rays (see also Section \ref{sec:sum} for possible contribution of the disk to X-rays).

In order to check the uncertainty of $\theta_F$, $\theta = 10^4$ are considered in Fig.\ref{fig:sed2}. Since the cases of $R = 10^{14}$ and $10^{15}$ cm were ruled out, only SEDs for $R = 10^{16}$ and $10^{17}$ cm are shown in the figure. Following the change of $\theta_F$, the spectral edge of synchrotron radiation is shifted to lower energy by a factor of $10^{2}$. Also, the maximum energy of $\gamma$-rays decreases, but this effect is smaller than the synchrotron edge because of the Klein-Nishina effect, which is the decrease of the IC cross-section compared with the Thomson scattering. The detectability of the $\gamma$-rays by CTA for $R = 10^{16}$ is not largely changed. Note that both Figs.\ref{fig:sed1} and \ref{fig:sed2} are consistent with the hard X-ray observations at present because the sensitivity of Swift BAT for 54 months is larger than the expected flux. Future observations in hard X-rays, e.g., NeXT \citep{Takahashi2008SPIE7011p18}, will constrain the maximum energy of electrons accelerated in the jets.

\begin{figure}
\begin{center}
\includegraphics[width=0.98\linewidth]{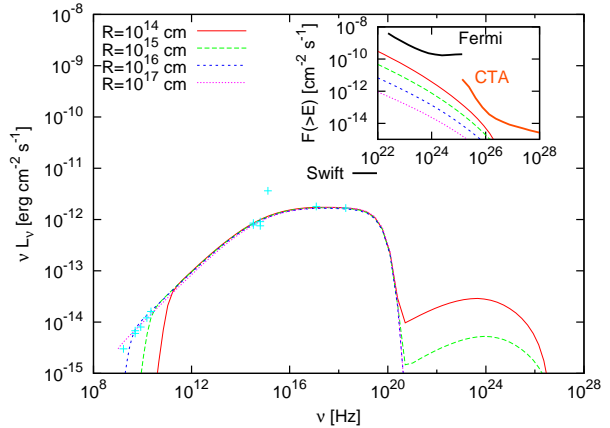} 
\caption{Same as Fig.\ref{fig:sed1}, but the cases of 
$\delta = 10$ and $\theta_F = 10^4$.}
\label{fig:sed3}
\end{center}
\end{figure}

\begin{table}
\begin{center}
\begin{tabular}{|c|c|c|c|} \hline
R [cm] & B [G] & $\eta$ \\ \hline \hline
$10^{14}$ & 5.86 & 0.25 \\
$10^{15}$ & 1.26 & 0.06  \\
$10^{16}$ & 0.272 & 0.014  \\
$10^{17}$ & 0.058 & 0.003  \\ \hline
\end{tabular}
\caption{Physical parameters adopted in Fig.\ref{fig:sed3} under $\delta = 10$.}
\label{tab:parameters2}
\end{center}
\end{table}

Finally, we discuss SEDs in the case of $\delta$ larger than 2.6. Fig.\ref{fig:sed3} shows SEDs calculated on the assumptions of $\delta = 10$ and $\theta_F = 10^4$. The other parameters are listed in Table \ref{tab:parameters2}. Since the high beaming factor boosts the flux of radiation, the energy density of radiation much lower than that in the case of Fig.\ref{fig:sed1} is enough to reproduce the observed flux up to X-rays. Thus, the energy density of magnetic field in the blob is dominated compared to that of radiation even for $R = 10^{14}$ cm, but the cases of $R \leq 10^{15.5}$ cm are not favored due to synchrotron self-absorption again. It is also dominated compared to the electron energy density, contrary to the cases of Figs.\ref{fig:sed1} and \ref{fig:sed2}. The smaller energy densities of radiation and electrons leads to much smaller ICS components than those in Fig.\ref{fig:sed1}. Thus, even long-term observations by Fermi LAT and CTA could not confirm these $\gamma$-rays.

\section{Discussion \& Summary} \label{sec:sum}

We have discussed possible $\gamma$-ray emission from jets in LLAGNs, and have also constrained several physical parameters in our SSC model. The size of an emission region (blob) was limited to $10^{16} \leq R \leq 10^{17.5}$ cm by a monthly time scale variability and synchrotron self-absorption. This range is consistent with the variability on typical time scale of a few days detected in LLAGNs \citep{Anderson2005ApJ627p674}. On the other hand, we have, so far, neglected a small ($\sim 10\%$) hours time scale ($\sim 1.5$ hr) variability in X-ray bands reported by \citet{Younes2010AA517p33}. Here, we discuss the origin of this variability.

This indicates the size of an emission region as 
\begin{equation}
R \leq \frac{\delta c \Delta t_{\rm obs}}{1 + z} 
= 5 \times 10^{14} \delta_3 \Delta t_{\rm obs,1.5h} ~~~{\rm cm}, 
\end{equation}
where $\Delta t_{\rm obs,1.5h} = \Delta t_{\rm obs} / 1.5$ hour is the time scale of the flux variability. This size is comparable with the Schwarzschild radius of the central black hole for $\delta = 2.6$ and is $\sim 2 \times 10^{15}$ cm even for $\delta = 10$. These sizes are not favored due to the limitation of $R$ if the X-rays are emitted from the blob. Thus, it is thought that X-ray radiation from an innermost part of an accretion disk contributes to this variability. Theoretically, ADAF models can predict X-rays via thermal bremsstrahlung of ions in an inner hot part of the disk (e.g., \citet{Manmoto1997ApJ489p791},\citet{Nemmen2010IAUS267p313}). This implies that emission from the disk could contribute to the total X-rays by a small (but significant) fraction, even if a jet component is dominated from radio to X-ray bands suggested by the correlation between radio and X-ray luminosities.

In this paper, a linear regime of synchrotron cooling of electrons, i.e., synchrotron cooling under constant magnetic field, was considered. Recently, non-linear radiative cooling of electrons has been discussed \citep{Schlickeiser2007AA476p1,Schlickeiser2008AA485p315}. Non-linear models assume a constant equipartition parameter $\eta^{-1}$ based on the success of spectral modeling of the observed blazars. Since the synchrotron and inverse compton cooling rates of electrons depend on the energy density of relativistic electrons, electron spectrum resultant from a solution of the kinetic equation of electrons is different from that derived from treatment in a linear regime even for a steady electron injection. Although whether equipartition is realized in jets of LLAGNs is not clear observationally at present, it is an intriguing topic to investigate common physical features between blazars (strong radio galaxies) and LLAGNs.

To summarize, we have discussed $\gamma$-ray emission from LLAGN jets in the framework of a SSC model on the assumption of radio and X-rays are dominantly produced from jets. The $\gamma$-rays are a direct probe of a jet component in radio to X-ray bands without contamination from the other components, although the predicted flux is not large. Several observational results allowed us to constrain physical parameters in the emission region in jets. In the case of a beaming factor as low as that of parsec scale jets and $R \sim 10^{16}$ cm, CTA may detect the $\gamma$-rays in the near future and test the jet domination of radiation from LLAGNs. The determination of the respective contribution of disk and jet components will gives us a hint of a physical connection between a disk and relativistic jet in LLAGNs. 

\section*{Acknowledgments}

H.T. thanks Yoshiyuki Inoue for useful discussions. H.T. thanks the referee for fruitful comments which helped to improve the manuscript. 

\bibliographystyle{mn.bst}
\bibliography{ms.bib}


\appendix
\onecolumn

\section{Synchrotron Self-Compton model} \label{app:ssc}

This appendix is dedicated to describe the detail of the synchrotron self-compton model used in this paper. We consider a steady state, that is a time-independent model. In addition, it is assumed that electrons and resultant synchrotron photons are distributed isotropically and magnetic field is randomly oriented in the blob. Below, the energies of electrons $E_e$ and photons $E_{\gamma}$ are measured in the unit of the electron mass, i.e., $\gamma = E_e / m_e c^2$ and $\epsilon_{\gamma} = E_{\gamma} / m_e c^2$.

The number density of photons in a blob frame, $d n_{\gamma} / d\epsilon$, is determined by a transport equation neglecting partial derivatives of time, 
\begin{equation}
\frac{1}{t_{\gamma,{\rm esc}}} 
\frac{d n_{\gamma}}{d \epsilon}(\epsilon) 
= \frac{d^2 n_{\gamma,{\rm syn}}}{dt d\epsilon}(\epsilon) 
+ \frac{d^2 n_{\gamma,{\rm ICS}}}{dt d\epsilon}(\epsilon), 
\end{equation}
where $t_{\gamma,{\rm esc}}$ is the escape time scale of photons from a blob, $d^2 n_{\gamma,{\rm syn}} / dt d\epsilon$ and $d^2 n_{\gamma,{\rm ICS}} / dt d\epsilon$ are the production rate of synchrotron photons and ICS photons per unit volume from the electron distribution assumed in Eq.\ref{eq:sp}. We assume an optically thin limit in this paper, that is $t_{\gamma,{\rm esc}} = R / c$.

The production rate of synchrotron photons is calculated from the production rate of synchrotron photons from an electron in random magnetic field \citep{Crusius1986AA164L16}, 
\begin{equation}
\frac{d^2 N_{\rm syn}}{dt d\epsilon} (\epsilon; \gamma) = 
\frac{3 \sqrt{3}}{\pi} 
\frac{\sigma_T c U_B}{\gamma \epsilon_B \epsilon m_e c^2} g^2 
\left[ K_{4/3}(g) K_{1/3}(g) - \frac{3}{5} g 
\left\{ {K_{4/3}}^2(g) - {K_{1/3}}^2(g) \right\} \right], 
\label{eq:trans}
\end{equation}
where
\begin{equation}
g = \frac{\epsilon}{3 \gamma^3 \epsilon_B}, ~~~
\epsilon_B = \frac{2 \pi \hbar c}{\gamma m_e c^2} 
\sqrt{\frac{2 r_e U_B}{\pi m_e c^2}}, 
\end{equation}
and $K_{p}(x)$ is the modified Bessel function of $p$ order. $r_e$ and $\hbar$ are the classical electron radius, Planck constant, respectively. Synchrotron radiation is accompanied by absorption in which a photon interacts with a charge in magnetic field. The absorption coefficient of synchrotron self-absorption is described in standard textbooks (e.g., \citet{Rybicki1979book}) as 
\begin{equation}
\alpha(\epsilon) = - \frac{1}{8 \pi c \epsilon} 
\left( \frac{2 \pi \hbar c}{m_e c^2} \right)^3 
\int_{\rm \gamma_{\rm min}}^{\rm \gamma_{\rm max}} d\gamma 
\frac{d^2 N_{\rm syn}}{dt d\epsilon} \gamma^2 
\frac{d}{d\gamma} \left[ \frac{1}{\gamma^2} 
\frac{d n_e}{d\gamma}(\gamma) \right]. 
\end{equation}
Thus, the production rate of synchrotron radiation is estimated as 
\begin{equation}
\frac{d^2 n_{\gamma,{\rm syn}}}{dt d\epsilon}(\epsilon) 
= \int d\gamma \frac{d n_e}{d\gamma}(\gamma) 
\frac{d^2 N_{\rm syn}}{dt d\epsilon} (\epsilon; \gamma) 
+ \frac{d n_{\gamma}}{d\epsilon}(\epsilon) \alpha(\epsilon) c. 
\label{eq:sync}
\end{equation}

The production rate of ICS is estimated as follows, 
\begin{equation}
\frac{d^2 n_{\rm ICS}}{dt d\epsilon}(\epsilon) 
= - c \frac{d^2 n_{\gamma}}{dt d\epsilon}(\epsilon) 
\int d\gamma \frac{d n_e}{d\gamma}(\gamma) R_{\rm ICS}(\gamma, \epsilon) 
+ c \int d\epsilon' \frac{d^2 n_{\gamma}}{dt d\epsilon}(\epsilon') 
\int d\gamma \frac{d n_e}{d\gamma}(\gamma) 
P_{\gamma, {\rm ICS}}(\epsilon; \gamma, \epsilon'). 
\label{eq:ics}
\end{equation}
where $R_{\rm ICS}(\gamma, \epsilon)$ and 
$P_{\rm ICS}(\epsilon; \gamma, \epsilon')$ are 
\begin{eqnarray}
R_{\rm ICS}(\gamma, \epsilon) = \int_{-1}^1 d\mu \frac{1 - \beta \mu}{2} 
\sigma_{\rm ICS} (\beta, \gamma, \epsilon), \\
P_{\gamma, {\rm ICS}}(\epsilon; \gamma, \epsilon') = \int_{-1}^1 d\mu \frac{1 - \beta \mu}{2} 
\frac{d \sigma_{\rm ICS}}{d\epsilon} (\epsilon; \gamma, \epsilon', \mu). 
\end{eqnarray}
$\mu$ is the cosine of the angle between the momentums of incident photons and electrons. $\sigma_{\rm ICS}$ is the cross-section of ICS in the case when an electron with the energy of $\gamma$ scatters a photon with the energy of $\epsilon$ \citep{Coppi1990MNRAS245p453}, 
\begin{equation}
\sigma_{\rm ICS}(\beta, \gamma, \epsilon) = \frac{3 \sigma_T}{4 x} 
\left[ \left( 1 - \frac{4}{x} -\frac{8}{x^2} \right) \log (1 + x) 
+ \frac{1}{2} + \frac{8}{x} - \frac{1}{2 (1 + x)^2} \right], 
\end{equation}
where $x = 2 \gamma \epsilon_{\gamma} ( 1 - \beta \mu )$. $d \sigma_{\rm ICS} / d\epsilon$ is the differential cross-section of ICS in the case when a photon with the energy of $\epsilon'$ is upscattered into the energy of $\epsilon$ by an electron with the energy of $\gamma$ \citep{Lee1998PhRvD58p043004}, 
\begin{equation}
\frac{d \sigma_{\rm ICS}}{d\epsilon} (\epsilon; \gamma, \epsilon', \mu) 
= \frac{3 \sigma_{\rm T}}{4 \gamma} \frac{1}{x} 
\left[ \frac{\gamma - \epsilon}{\gamma} 
+ \frac{\gamma}{\gamma - \epsilon} 
- \frac{4}{x} \left( \frac{\epsilon}{\gamma - \epsilon} \right) 
+ \frac{4}{x^2} \left( \frac{\epsilon}{\gamma - \epsilon} \right)^2 \right]. 
\end{equation}
Thus, $R_{\rm ICS}(\gamma, \epsilon)$ and $P_{\gamma, {\rm ICS}}(\epsilon; \gamma, \epsilon')$ can be interpreted as angle-averaged cross-sections.

After Eqs.\ref{eq:sync} and \ref{eq:ics} are substituted, Eq.\ref{eq:trans} is an integral equation categorized to the Fredholm equation of the second kind. This equation can be numerically solved by the Nystrom method (e.g., \citet{NRC1992book}). Note that $d n_{\gamma} / d\epsilon$ includes higher order ICS photons, whose seed photons are ICS photons. Finally, photon flux observed at Earth is estimated as 
\begin{equation}
\frac{d^2 N_{\gamma}}{dt_{\rm obs} d\epsilon_{\rm obs}}(\epsilon_{\rm obs}) 
= \delta^2 \frac{V}{4 \pi {d_L}^2 t_{\gamma,{\rm esc}}} 
\frac{d n_{\gamma}}{d\epsilon}\left( \frac{( 1 + z ) \epsilon_{\rm obs}}{\delta} \right), 
\end{equation}
where $\epsilon_{\rm obs} = ( 1 + z )^{-1} \delta \epsilon$ and $d_L$ is the luminosity distance of the source. 

\label{lastpage}

\end{document}